\documentclass[letterpaper,10pt]{article}
\usepackage{osameet2}
\usepackage{verbatim}
\usepackage{amsmath,amssymb}
\usepackage[font=scriptsize]{subcaption}
\usepackage[font=scriptsize]{caption}

\begin{document}

\title{Optical Phase Imaging Using Synthetic Aperture Illumination and Phase Retrieval}

\author{Dennis J. Lee and Andrew M. Weiner}
\address{School of Electrical and Computer Engineering,
    Purdue University, West Lafayette, 47906 IN, USA}
\email{leedj@purdue.edu}

\begin{abstract}
We perform quantitative phase imaging using phase retrieval
to implement synthetic aperture imaging.  Compared to
digital holography, the developed
technique is simpler, less expensive, and more
stable.
\end{abstract}


Optical phase imaging finds important applications in
biomedical imaging where samples are often transparent and
weakly scattering. Once the phase and complex field are known, label-free cell imaging, numerical refocusing, and
differential interference contrast can be performed
\cite{Kim12, Marquet05, Rappaz05}.

An important concern in any imaging system is resolution.
The technique called \emph{synthetic aperture imaging}
illuminates a sample at multiple angles and measures a
complex field image at each angle.  Conceptually, each
angular field covers a different portion of frequency space.  Effectively, the numerical
aperture is increased.  Currently this technique is
implemented using digital holography.  Although digital holography is commonly used, it does have
its disadvantages.  
The reference arm adds
more parts, cost, and complexity.  For example, the phase shifting
interferometry in \cite{Kim12} requires AOMs to upshift the
reference beam and a high frame rate camera to
capture the phase-shifted images.

We propose to implement synthetic aperture imaging using
phase retrieval.  A separate reference beam
is not required, which aids stability.  The measurements are based on defocused
intensity images, which does not require an expensive high
frame rate camera.  Fewer components are required, which
\begin{figure}[htbp]
\begin{minipage}{.79\textwidth}
    \centering
    \includegraphics[width=0.5\linewidth]{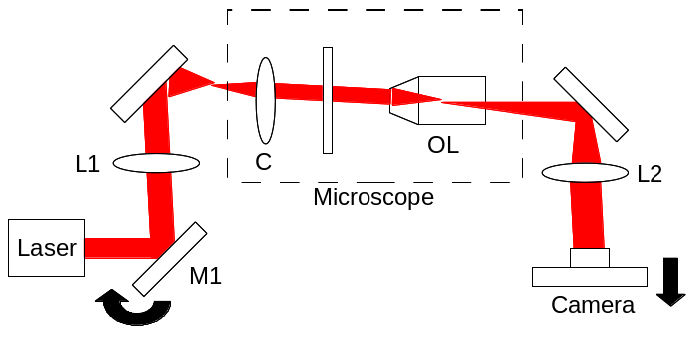}
    \caption{Experimental setup for synthetic aperture phase
retrieval.  Laser: He-Ne, $ \lambda = 633 $ nm; M1: gimbal mount mirror; \\
L1: lens (f = 300 mm);
C: condenser lens (NA 1.4); OL: objective lens (NA 0.75);
L2: tube lens (f = 200 mm).}
    \label{fig:f_setup}
\end{minipage}
\begin{minipage}{.2\textwidth}
        \centering
        \begin{subfigure}{.5\textwidth}
            \centering
            \includegraphics[width=\linewidth]{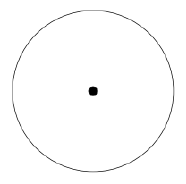}
            \caption{Case 1.}
            \label{fig:fa_circles}
        \end{subfigure}%
        \begin{subfigure}{.5\textwidth}
            \centering
            \includegraphics[width=\linewidth]{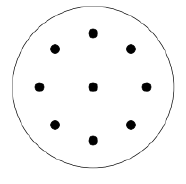}
            \caption{Case 2.}
            \label{fig:fb_circles}
        \end{subfigure}
    \caption{Angular scanning at back focal plane of
condenser.}
    \label{fig:f_circles}
\end{minipage}
\vspace{-\baselineskip}
\end{figure}
reduces cost.  Our setup is shown in Fig. \ref{fig:f_setup}.
Mirror M1 is a motorized gimbal mount which steers the beam
at different angles to provide oblique illumination at the
sample.  The sample is imaged on a camera which is mounted
on a translation stage to defocus the sample.  

We process these defocused images $ I_1,
..., I_N $ using phase retrieval based on an iterative
algorithm \cite{Pedrini05, Almoro06}.  We start
in the first plane with complex amplitude $ U_1 = \sqrt{I_1}
$ and phase $
\phi_1(x,y) = 0 $.  We numerically propagate the complex
amplitude $ U_{n-1} $ at plane $ (n-1) $ to plane $ n $ and extract phase $
\phi_n(x,y) $.  We update the complex amplitude as $ U_n =
\sqrt{I_n} \text{exp}[ j \phi_n(x,y) ] $.  This propagate
and update procedure is repeated as many times as necessary.
Convergence is checked by numerically propagating the
retrieved field and comparing with measured intensities.  For each angle of illumination, we apply this
procedure to calculate phase and the complex field.  We sum the complex fields at each
angle to calculate the synthesized field.  The
\emph{synthesized phase} is the phase of the synthesized
field.  In our experiment, for each angle, we measure 11
intensity images at planes separated by 2.1 $\mu m$,
symmetric about the focal plane at $ z = 0 $.  The samples
are 10 $ \mu m $ polystyrene beads (n = 1.587) immersed in oil (n =
1.515).  

The simplest test is to measure a single angle at
$0^{\circ}$
\begin{figure}
\centering
\begin{subfigure}{.30\textwidth}
    \centering
    \includegraphics[width=\linewidth]{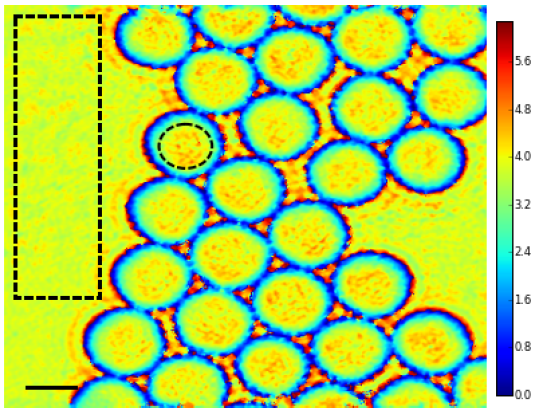}
    \caption{Wrapped phase (rad).}
\end{subfigure}%
\begin{subfigure}{.30\textwidth}
    \centering
    \includegraphics[width=\linewidth]{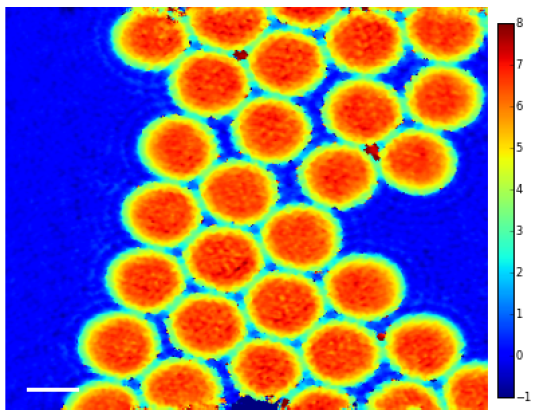}
    \caption{Unwrapped phase (rad).}
\end{subfigure}
\vspace{-\baselineskip}
\caption{Phase images at 0 degrees. Scale bar: 6 $
\mu m $.}
\label{fig:f_dc}
\vspace{-\baselineskip}
\end{figure}
 illumination (Fig. \ref{fig:fa_circles}).  Figure
\ref{fig:f_dc} shows the resulting phase image, which agrees
with refractive index calculations.
The wrapped
phase image most clearly
exhibits diffraction noise, evidenced by the diffraction
\begin{figure}[htbp]
\centering
\begin{minipage}{.30\textwidth}
    \centering
    \includegraphics[width=\linewidth]{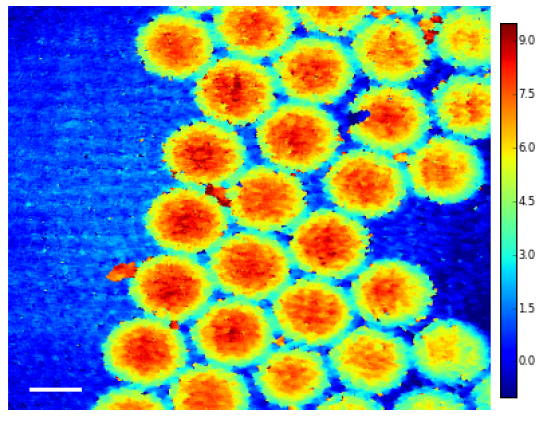}
    \caption{Interferometry: Unwrapped phase (rad).}
    \label{fig:f_offaxis}
\end{minipage}
\begin{minipage}{.58\textwidth}
        \centering
        \begin{subfigure}{.5\textwidth}
            \centering
            \includegraphics[width=\linewidth]{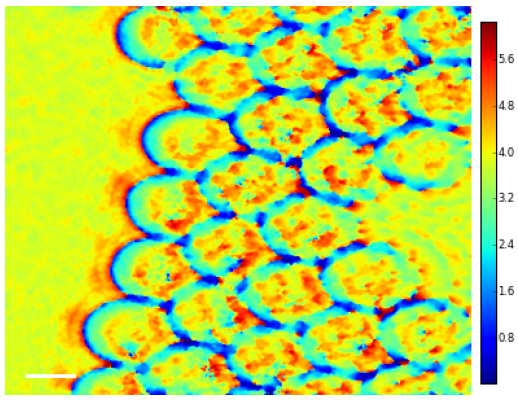}
            \caption{Wrapped phase (rad).}
        \end{subfigure}%
        \begin{subfigure}{.5\textwidth}
            \centering
            \includegraphics[width=\linewidth]{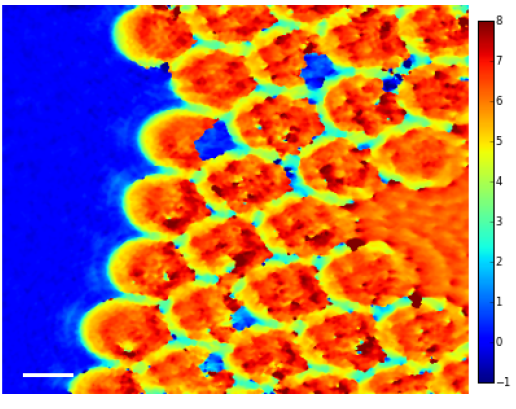}
            \caption{Unwrapped phase (rad).}
        \end{subfigure}
    \vspace{-\baselineskip}
    \caption{Phase images at 12.3 degrees.  Scale bar: 6 $
\mu m $.}
    \label{fig:f_phase_1_1}
\end{minipage}
\vspace{-2\baselineskip}
\end{figure}
rings and speckle-like patterns inside the beads.  For
reference, we also measure phase with off-axis
interferometry (Fig. \ref{fig:f_offaxis}), which shows
approximately the same phase shift at the center of the
beads.  

\begin{figure}[h]
\centering
\begin{subfigure}{.30\textwidth}
    \centering
    \includegraphics[width=\linewidth]{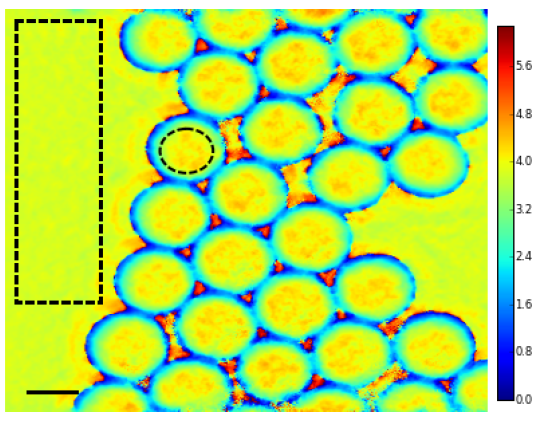}
    \caption{Wrapped phase (rad).}
\end{subfigure}%
\begin{subfigure}{.30\textwidth}
    \centering
    \includegraphics[width=\linewidth]{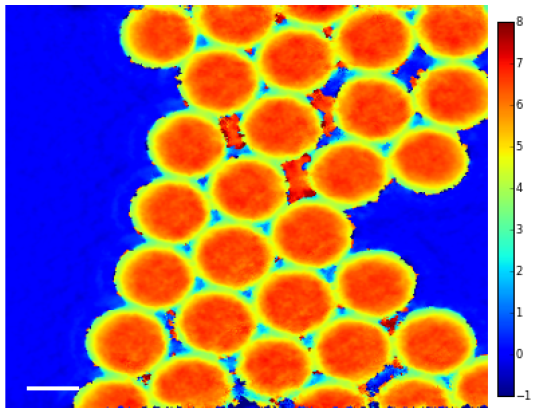}
    \caption{Unwrapped phase (rad).}
\end{subfigure}
\vspace{-\baselineskip}
\caption{Synthesized phase images. Scale bar: 6
$ \mu m $.}
\label{fig:f_synthesized_8}
\vspace{-2\baselineskip}
\end{figure}

In test case 2, we measure 9 total angles by scanning the beam at
 the back focal plane of the condenser
(Fig. \ref{fig:fb_circles}).  The beam is scanned in an
approximate circle pattern so that the largest angle of
illumination is $12.3^{\circ}$ at the sample.  Note the
angular spectra overlap between angles.  For each
angle we measure a phase image.  Figure
\ref{fig:f_phase_1_1} shows a phase image at $12.3^{\circ}$.
Although
there are unwrapping errors,
the figure illustrates the basic idea.  Since the beam is oblique
to the camera, the beads appear elongated \cite{Kim12,
Choi07}.  To the best of our knowledge, this work is the
first experimental demonstration of phase retrieval on an
obliquely illuminated sample.  Next we add the measured
complex fields at each angle to form the synthesized field,
with the background phases set equal.
Figure \ref{fig:f_synthesized_8} shows the synthesized
phase.  Compared to test case 1 (Fig. \ref{fig:f_dc}), the phase image displays
reduced diffraction noise; the diffraction rings and phase
speckle features inside the beads are mitigated.  We measure
the phase variance inside the bead (dashed circle)
for Fig. \ref{fig:f_dc} ($ \sigma^2 = 0.118)  $ and for Fig.
\ref{fig:f_synthesized_8} ($ \sigma^2 = 0.047 $); the phase
speckle noise reduced by 60\%.  We also measure the
background variance (dashed rectangle) for Fig.
\ref{fig:f_dc} ($ \sigma^2 = 0.023 $) and for Fig.
\ref{fig:f_synthesized_8} ($ \sigma^2 = 0.004 $); the
background noise reduced by 83\%.  

We have demonstrated the principle of using phase retrieval to
implement synthetic aperture imaging.  Our experiment used a
0.75 NA objective; it can be extended to the 1.4 NA used in
\cite{Kim12}.  Our approach enables resolution enhancement
without needing an expensive high NA objective.  Overall this new imaging
technique is simpler, less expensive, and more stable than
digital holography, and it paves the way for other applications such as
tomographic phase microscopy to be enabled by phase
retrieval.

\end{document}